\documentclass[11pt,twoside]{article}

\usepackage{asp2006}
\usepackage{graphicx}

\begin{document}
\title{Secular evolution of disk galaxies}  
\author{F. Combes}  
\affil{Observatoire de Paris, LERMA, 61 Av. de l'Observatoire, F-75014, Paris, France}   

\begin{abstract} 
Galaxy disks evolve through angular momentum transfers
between sub-components, like gas, stars, or dark matter halos,
through non axi-symmetric instabilities.
The speed of this evolution is boosted in presence
of a large fraction of cold and dissipative gas component.
When the visible matter dominates over the whole disk,
angular momentum is exchanged between gas and stars only.
The gas is driven towards the center by bars, stalled transiently
in resonance rings, and driven further by embedded bars,
which it contributes to destroy. From a small-scale molecular
torus, the gas can then inflow from viscous torques, dynamical
friction, or $m=1$ perturbations.
 In the weakened bar phases, multiple-speed spiral patterns can
 develop and help the galaxy to accrete external gas flowing
from cosmic filaments.  The various phases of secular evolution
are illustrated by numerical simulations.
\end{abstract}

\vspace{-0.5cm}
\section{Bars, gas flows and secular evolution}

\subsection{Bar formation and evolution}

Non-axisymmetries and in particular bars are
the motor of secular evolution, in transferring
the angular momentum in galaxies.  In the 80' and 90',
numerical simulations established how bars formed,
in pure stellar disks, without any dark matter haloes,
or embedded in rigid halo components: the angular momentum
was exchanged within the disk, the outer parts
gaining the momentum from the inner parts.
This exchange occured mainly at bar formation, the
bar pattern speed slowing down slightly, while the stellar
orbits in the bar were more and more elongated,
corresponding to a lower precessing rate.
Once formed, the bar was then robust (e.g. Sellwood 1981).

Already, some dynamical heating was observed to
produce some feedback in the bar strength evolution:
more unstable stellar disks develop a bar sooner, but
then end up with a weaker bar than more stable disks
(Combes et al 1990).

\subsection{Bars with gas disks}

The introduction of gas in disks completely changes
the stability, due to dissipation.
The angular momentum can be exchanged between gas and stars,
and from inner to outer parts,
again with a negligible halo component.
While gas flows towards the center, the bar is destroyed and
gives rise to a triaxial bulge (Friedli \& Benz 1993).

All this evolution can be traced back to the
bar gravity torques acting on the gas. The torques
are proportional to the
phase shift between the gas and the potential wells
due mainly to the stars in the bar. Dissipation produces
these phase shifts, which are conspicuous through the
characteristic leading dust lanes in barred galaxies.
The gas then loses angular momentum
and mass concentrates towards the center.

The amplitude of the phenomenon has been 
quantified from observations of barred galaxies.
The gravity torques can be computed from the
stellar potential deduced from the red image of
the galaxy, and the effective angular momentum
exchange computed from the gas distribution,
obtained through H$\alpha$, HI or CO emission images.
 In strongly barred galaxies, the gas inside corotation
inflows to the center in one or two rotations.
Even in weakly barred galaxies, the gravity torques
are efficiently producing gas flows, and fueling the nucleus,
as for instance in NGC 6574 (Lindt-Krieg et al 2007)
or NGC 3147 (Casasola et al 2008).

\subsection{Formation of rings}

The secular evolution as described above, 
the gas inflow driven by bars, finds
its confirmation in the frequent observation of resonant rings,
where gas is piling up, and form stars in bright knots.
The gravity torques change sign at each
resonance, and cancel in the rings where the gas disctribution is
symmetrically distributed with respect to the stars.

Galaxies often possess multiple-rings,
corresponding to the various
Lindblad resonances, ILR, UHR, OLR (Buta \& Combes 1996).
 With only one bar, the gas stops its inflow at the ILR, in 
the nuclear ring.  The decoupling of a nuclear bar inside the 
nuclear ring triggers the AGN fueling (cf NGC 2782, Fig 1).

\begin{figure}
\begin{center}
\includegraphics[angle=-90,width=12cm]{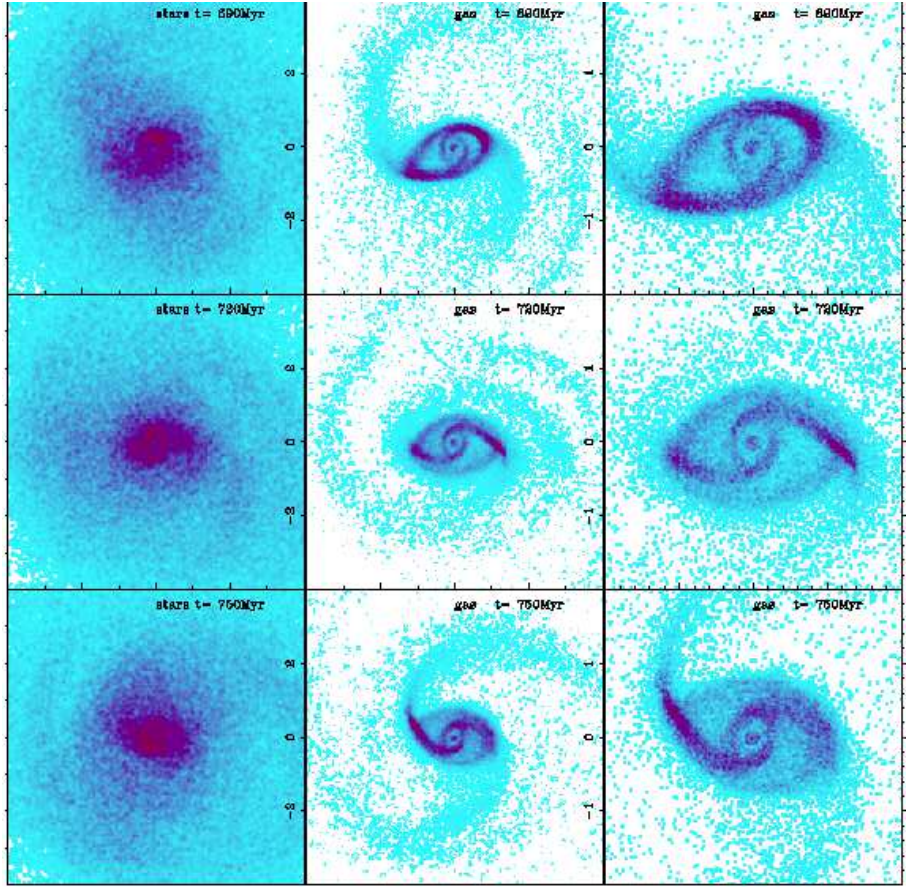}
\end{center}
\caption{ Gray scale plot of the stellar component
({\it left}), the gas component ({\it middle}), and an expanded
version of the gas distribution  ({\it right}), in a self-consistent  simulation
of a double bar formation, meant to reproduce the galaxy NGC 2782
(the axes are in kpc). It can be seen that the gas, which was predominantly
in the more external ring, corresponding to the ILR of the primary bar, at T=690 Myr,
is falling progressively inward, and is found inside the ILR at T=750 Myr.
From Hunt et al (2008).}
\label{n2782}
\end{figure}

\vspace{-0.3cm}
\section{Angular momentum transfer with haloes}

When a massive dark halo is taken into account around
spiral galaxies, the angular momentum transfer occurs mainy 
from baryons to the dark matter halo
(Athanassoula 2002, 2003).
This dominates over the exchange with gas, or
between the inner and the outer disk. This loss
of angular momentum boosts the bar strength.
 In addition, due to the bar dynamical friction
against the dark matter particles in the halo, the 
bar pattern speed decreases regularly
(Debattista \& Sellwood 2000). The low value obtained at the end 
is not compatible with the high pattern speeds observed. This
constrains the amount of dark matter in the inner parts
of the galaxy, and favors a maximum disk.

A consequence of the angular momentum exchange,
is the re-distribution of matter across the disk,
which could be at the origin of 
surface density breaks, frequently observed in
spiral galaxies (e.g. Pohlen, 2002). For instance
the outer disk breaks could correspond to
the outer Lindblad resonance (Debattista et al 2006).
Other mechanisms have been invoked for
the existence of these breaks (Roskar et al 2008, Figure 2).
When considering the inside-out formation of a disk galaxy from a halo
of gas, the break corresponds to a star formation cut-off, due
to a threshold in gas density,
and the exponential radial distribution beyond the break is due
to scattering of stars by spiral arms. 

\begin{figure}
\begin{center}
\includegraphics[width=12cm]{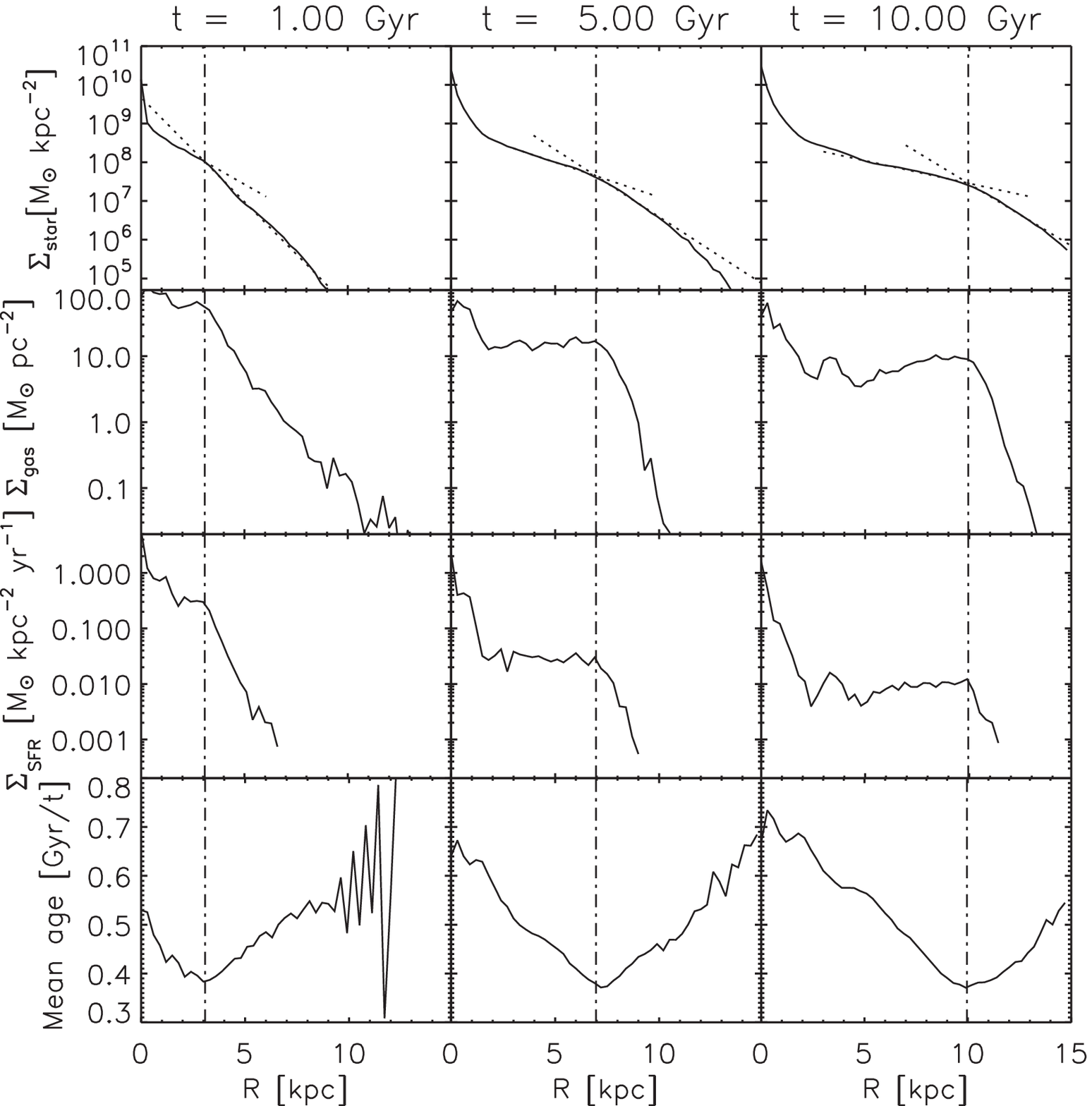}
\end{center}
\caption{Radial distributions at various epochs (1, 5, and 10 Gyr)
of the stellar surface density, the gas surface density,
the star formation rate, and the  mean stellar age, of
a disk simulation by Roskar et al (2008). The stellar
density is fitted by two exponentials, which intersection 
 is taken as the break radius (materialized by the
vertical lines in all panels).}
\label{roskar}
\end{figure}

If the bar strengthens through angular momentum transfer
to the dark matter halo, it is choked by
the vertical resonance and peanut formation.
During the standard bar formation, the pattern speed
$\Omega_b$ increases first, due to mass concentration,
 and decreases slightly after.
 In presence of a massive halo, the pattern speed slows down 
 much more, and the resonance moves outward in radius.
There is then two successive bucklings
(Martinez-Valpuesta et al 2006).

\vspace{-0.3cm}
\section{Bar destruction,  formation of lenses}

\subsection{Mechanisms of destruction}

It is known from the first self-consistent numerical simulations
of disks with gas that bars can be destroyed (Friedli \& Benz 1993).
 This destruction was first thought to be due only 
to the central mass concentration (CMC), produced by the gas inflow
(Norman et al 1996).
The chaos induced by the superposition of a CMC and a bar
destroys bars in pure stellar disks.
For pure stellar models, a CMC is more
efficient to destroy bars, in disk-dominated models.
The bar pattern speed increases when
bar weakens, which is the reverse when bar strengthens
(Athanassoula et al 2005).

When the gas fraction in a spiral galaxy
is larger than 5\% of the total mass, then
the gas inflow alone is able to destroy the bar
(Bournaud \& Combes 2002).
Gas is driven inwards by the bar torques.
The angular momentum lost by the gas is taken up 
by the bar wave, and this destroys the bar,
which is a wave with negative momentum inside corotation.
The main destruction mechanism is therefore the torque
exerted from the gas to the bar, and 
not only the presence of the Central Mass Concentration.
A CMC of only 1\% is not sufficient to destroy the bar
(Shen \& Sellwood 2004).
But 1-2\% of gas infall is enough to transform a bar in a lens
(Friedli 1994, Berentzen et al 1998, Bournaud et al 2005a).
The final effect of gas depends on the cooling:
the bar is destroyed by an isothermal but not 
an adiabatic gas (Debattista et al 2006).

\subsection{Reformation of bars}

For all gas-rich intermediate and late-type spirals,
the observed frequency of bars means that bars have
to be reformed after their destruction. If
external gas is accreted through the outer parts, the
secular evolution can involve a self-regulated cycle,
where the bar is destroyed by the gas inflow it
has produced, but can reform when the gas disk
is replenished by external accretion.
 When the bar is strong, accreted gas is stalled 
at the OLR, since positive gravity torques are
exerted between CR and OLR. The gas inflow is intermittent.
When the bar weakens, the gas can replenish the disk, to make
 it unstable again to bar formation.

In a recent work, Berentzen et al (2007) have
found on the contrary that the angular momentum
exchange between the gas and the stars was quite low,
and that most of the exchange was between the stars
and the dark matter halo. Their gas fraction
was not high (f$_{gas}$ $<$ 8\%), but their disks were
dominated by the dark halo.
In spite of this, the bar is destroyed more quickly in the presence
of gas, and the destruction efficiency is
correlated with the gas fraction.
Their main bar destruction is through a peanut formation,
however the presence of gas prevents peanuts.
The bar destruction is attributed to the CMC.
The fraction of gas able to destroy bars is higher
in presence of massive dark matter haloes.
When the DM halo mass is negligible within the disk, the gas fraction
of 6\% is sufficient. 
It becomes 20\%, when the DM/disk mass ratio is 3 (Curir et al 2007).

The influence of the dark matter halo is well
studied through the comparison with galaxy models
in modified gravity. Numerical simulations have been done
with identical disks as initial conditions, with the same rotation curve. 
 The latter is flat, in one case due to a surrounding
dark matter halo, in the other case due to the MOND dynamics 
(Tiret \& Combes 2007). 
The bar strength and pattern speed are quite
different with and without DM.

With DM, the bar appears later, and can reform 
after the peanut-driven bar weakening through angular momentum exchange
with the halo, but the pattern speed $\Omega_b$ falls off.
The peanut moves outwards in radius
due to the slowing down of the bar.
With MOND, disks are more unstable, and stronger bars form
earlier, they are also weakened by peanut formation, but later.
They cannot reform afterwards, in the absence of halo particles
to absorb the angular momentum. Overall on a series of models representing
the Hubble sequence, bars are more frequent in MOND.

\subsection{Are lenses dissolved bars ?}

The suggestion has been made by Kormendy (1979) that lenses
could be the signal of destroyed bars.  In particular, the sharp edge
of lenses coincides with a potential resonance:
inner lenses extend to what was an inner ring, outer lenses extend
to the old outer ring. At the extremities of weak bars,
are frequently observed "ansae", or bright spots ending the oval/bar.
The frequency of "ansae" is strong for early-type galaxies, and then decreases
significantly towards the late-types (Martinez-Valpuesta et al 2007).

Lenses are stellar components of nearly constant surface density,
with a sharp edge, and steep outer gradient, as shown by the prototypical
examples N1553 (Freeman 1975) or N1291 (Kormendy 1982).
They are as flat as disks, and have
the same sizes and colors than bars.
Their kinematics reveal high velocity dispersion in the center, and low
in the outer parts.  54\% of SB0-SBa have lenses,
but not late-type nor early-type unbarred galaxies (Kormendy 1979).
The bar fills the lens in one dimension, mimicking a $\Theta$-like 
morphology. There is therefore no doubt that lenses are associated to bars.

\begin{figure}
\begin{center}
\includegraphics[angle=-0,width=10cm]{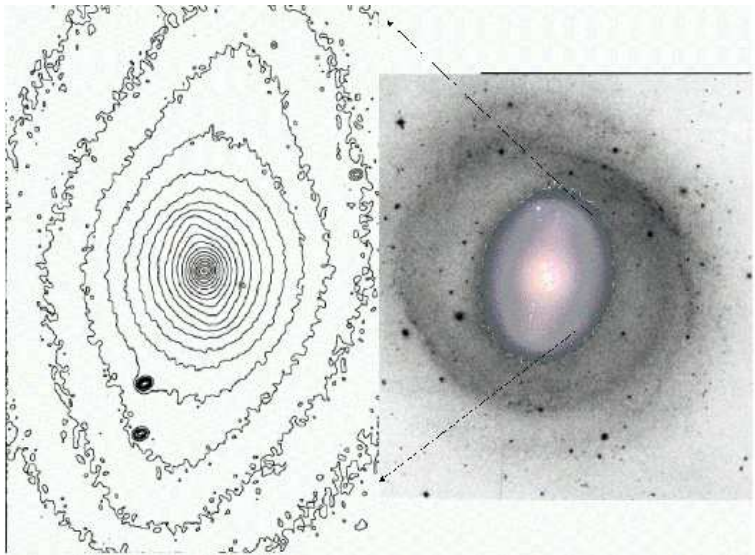}
\end{center}
\caption{ Visible image of NGC 1291 ({\it right}), and an expanded
version of the central distribution ({\it right}), in the near-infrared.}
\label{n1291}
\end{figure}

Numerical simulations of bar destruction by a CMC are
compatible with the hypothesis of lenses being destroyed bars. 
The bar is weakened when 
chaotic orbits dominate in the region between x2 orbits near the CMC
and 4/1 orbits near corotation.
In the chaotic sea, orbits are bounded only inside corotation, which could
explain the sharp cut-off of the lens;
the remaining regular orbits near CR would provide the "ansae" morphology. 
NGC 1291, a lens prototype, is probably experiencing a destruction
phase: it has an embedded bar, as shown in Figure 3.
It is an SB0a spiral, with blue, HI gas-rich, outer ring (OLR), with
hot X-ray gas in the center, of low metallicity (0.1Z$_\odot$).
This low-abundance gas could be due to infall (Perez \& Freeman 2006).

\vspace{-0.3cm}
\section{Lopsidedness (m=1 modes)}

Secular evolution involves also asymmetrical
modes, such as lopsidedness. About half of all galaxies
possess an asymmetry in their HI distribution 
(Richter \& Sancisi 1994). This fraction increases
to 77\% for late-types (Matthews et al 1998).
About 20\% of galaxies have their $m=1$ Fourier
coefficient in stellar density A$_1>$0.2, from their 
near-infrared images.
 This high degree of asymmetry could be driven by
a dark matter halo asymmetry (Angiras et al 2007).
Some could be due to companions, but it is not the
general case, since the $m=1$ perturbation is then
short-lived. Asymmetric gas accretion could 
provide a mechanism (Bournaud et al 2005b).
Alternatively, counter-rotating disks produce lopsidedness
(Dury et al, this conference).

It is also possible that global $m=1$ modes develop
in stellar disks, without any dark matter halo. These
are modes of long wavelength, of the order of the disk radius
(Saha et al 2007, Figure 4). Their precession rate is low, and
the self-gravitating modes have a small pattern speed.

\begin{figure}
\begin{center}
\includegraphics[angle=-90,width=10cm]{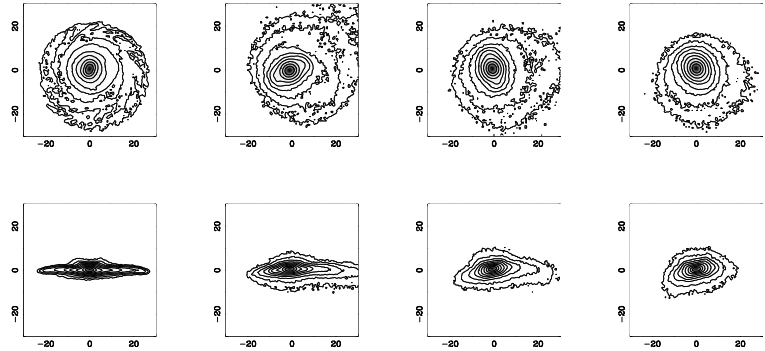}
\end{center}
\caption{Simulations of the $m=1$ instability
in a purely stellar disk, without dark halo. 
Contours in logarithmic scale of the surface density of
the stellar disc, face-on (top) and edge-on (bottom), at four different
epochs: T=0.8, 5.6, 9.6, 14.4 Gyr, from left to right. 
From Saha et al (2007).}
\label{kanak}
\end{figure}

\vspace{-0.3cm}
\section{Conclusion}
\vspace{-0.2cm}

Secular evolution is important for bars, it
controls their strength, their pattern speed, 
or their vertical thickness through peanut formation.
The dominant angular momentum transfer from the disk occurs
with the DM halo or with the outer disk,
according to the halo to disk mass ratio, or between the
stellar and gaseous components, according to the dissipation character
of the gas.

Bars are weakened by central mass concentrations and/or gas inflows,
driven by bars themselves, implying self-regulation.
The gas fraction able to destroy bars depends on the dark matter 
halo to disk mass ratio.
Lopsidedness ($m=1$ global mode) can also weaken the bar.
Bars dissolve into lenses, especially in early-type galaxies.

%\acknowledgements %%% Text of acknowledgements runs on after this command.

%%% THE BIBLIOGRAPHY


\vspace{-0.3cm}
\begin{thebibliography}{}
\vspace{-0.2cm}
\bibitem[]{} Angiras, R.A., Jog, C.J., Dwarakanath, K.S., Verheijen, M.:2007 MNRAS 378, 276
\bibitem[]{} Athanassoula, E.: 2002, ApJ 569, L83
\bibitem[]{} Athanassoula, E.: 2003, MNRAS 341, 1179
\bibitem[]{} Athanassoula, E., Lambert, J. C., Dehnen, W.: 2005, MNRAS 363, 496
\bibitem[]{} Berentzen I., Heller, C. H., Shlosman, I., Fricke, K. J.: 1998, MNRAS 300, 49
\bibitem[]{} Berentzen I., Shlosman, I., Martinez-Valpuesta, I., Heller, C.: 2007, ApJ 666, 189
\bibitem[]{} Bournaud F., Combes F.: 2002, A\&A 392, 83
\bibitem[]{} Bournaud F., Combes F., Semelin B.: 2005a, MNRAS 364, L18
\bibitem[]{} Bournaud, F., Combes, F., Jog, C. J., Puerari, I. : 2005b, A\&A  438, 507
\bibitem[]{} Buta, R. \& Combes, F. 1996, Fundamentals of Cosmic Physics, Volume 17, pp. 95-281
\bibitem[]{} Casasola V., Combes F., Gardia-Burillo S. et al: 2007, A\&A submitted
\bibitem[]{} Combes F., Debbasch F., Friedli D., Pfenniger D.: 1990, A\&A  233, 82
\bibitem[]{} Curir, A., Mazzei, P., Murante, G.:  2007 A\&A   467, 509
\bibitem[]{} Debattista V.P., Sellwood J.: 2000, ApJ 543, 704
\bibitem[]{} Debattista V.P., Mayer, L., Carollo, C. M. et al.: 2006, ApJ 645, 209
\bibitem[]{} Dury V., de Rijcke S., Debattista V., Dejonghe H.: 2008, in ``Formation and Evolution of Galaxy Disks'',
                          ed. J. Funes and E. Corsini
\bibitem[]{} Freeman K.: 1975  in ``Dynamics of stellar systems'',  ed. A. Hayli, Kluwer, p. 367 % N1553
\bibitem[]{} Friedli, D., Benz W.: 1993, A\&A 268, 65  % destruction of bars
\bibitem[]{} Friedli, D., 1994, in Mass-Transfer Induced Activity in Galaxies, Ed I. Shlosman, 
    Cambridge University Press,  p.268
\bibitem[]{} Hunt L., Combes F., Garcia-Burillo S. et al 2008, A\&A, submitted
\bibitem[]{} Kormendy J.: 1979, ApJ 227, 714
\bibitem[]{} Kormendy J.: 1982, in Morphology and dynamics of galaxies; Saas-Fee, p. 113-288.
\bibitem[]{} Lindt-Krieg E., Eckart A., Neri R., et al: 2007, A\&A in press (arXiv:0712.3133) 
\bibitem[]{} Martinez-Valpuesta I., Shlosman, I., Heller, C.: 2006 ApJ 637, 214
\bibitem[]{} Martinez-Valpuesta, I., Knapen, J. H., Buta, R.: 2007 AJ 134, 1863  % ansae
\bibitem[]{} Matthews, L. D., van Driel, W.,  Gallagher, J. S.: 1998, AJ, 116, 2196
\bibitem[]{} Norman C., Sellwood J.A., Hasan H.: 1996 ApJ 462, 114
\bibitem[]{} Perez I., Freeman K.: 2006 A\&A 454, 165
\bibitem[]{} Pohlen M.: 2002,  PhD Thesis, Ruhr-Universit\"at, Bochum, Germany
\bibitem[]{} Richter, O. -G.,  Sancisi, R. 1994, A\&A, 290, L9
\bibitem[]{} Roskar R., Debattista V.P., Stinson G.S. et al: 2007, ApJL in press (arXiv:0710.5523) 
\bibitem[]{} Saha K., Combes F., Jog C.J.: 2007, MNRAS 382, 419
\bibitem[]{} Sellwood, J.A: 1981 A\&A 99, 362
\bibitem[]{} Shen J., Sellwood J.A.: 2004 ApJ  604,  614
\bibitem[]{} Tiret, O., Combes, F.: 2007, A\&A 464, 517

\end{thebibliography}
\end{document}